\documentclass[aps,prl,a4paper,twocolumn,floatfix,showpacs,preprintnumbers]{revtex4-1}
\usepackage{amsmath}
\usepackage{amssymb}
\usepackage{graphicx}
\usepackage{color}
\usepackage{epstopdf}

\begin{document}

\title{Observation of quantum states without a semiclassical equivalence bound by a magnetic field gradient}
\author{B. Sch\"uler, M. Cerchez, Hengyi Xu, J. Schluck and T. Heinzel}
\affiliation{Condensed Matter Physics Laboratory, Heinrich-Heine-Universit\"at, 40225 D\"usseldorf, Germany}
\author{D. Reuter}
\affiliation{Department Physik, Universit\"at Paderborn, 33098 Paderborn, Germany}
\author{A. D. Wieck}
\affiliation{Lehrstuhl f\"ur Angewandte Festk\"orperphysik, Ruhr-Universit\"at Bochum, 44780 Bochum, Germany
}

\date{\today }

\begin{abstract}
Resonant transmission through electronic quantum states that exist at the zero points of a magnetic field gradient inside a ballistic quantum wire is reported. Since the semiclassical motion along such a line of zero magnetic field takes place in form of unidirectional snake trajectories, these states have no classical equivalence. The existence of such quantum states has been predicted more than a decade ago by theoretical considerations of Reijniers and coworkers [\onlinecite{Reijniers2002}]. We further show how their properties depend on the amplitude of the magnetic field profile as well as on the Fermi energy.
\end{abstract}

\maketitle

Magnetic field gradients are ubiquitous in science and technology. They are key components of functional units as diverse as, e.g., the Stern-Gerlach setup,\cite{Gerlach1922} Tokamaks for plasma confinement,\cite{Sakharov1958,Tamm1958} magnetic resonance imaging,\cite{Lauterbur1973,Garroway1974} diamagnetic levitation,\cite{Berry1997} para-hydrogen production,\cite{Juarez2002} or read/write heads of computer hard disks based on the giant magnetoresistance.\cite{Baibich1988,Binasch1989} In these applications, the magnetic gradient either acts on the electron or nuclear spin, or it affects the dynamics of charged particles, which can be understood within classical pictures. Orbital quantization effects in inhomogeneous magnetic fields have attracted much less attention but may become relevant in some devices as their downsizing continues.

Low-dimensional electron gases in semiconductors are excellent systems for studying such effects, since the electron Fermi energy can be comparable to the magnetic confinement energy, characterized by $\hbar\omega_c$, where $\omega_c$ denotes the cyclotron frequency. Furthermore, the electron gases can be exposed to strong magnetic gradients by ferromagnetic \cite{Ye1995} or superconductive \cite{Carmona1995} electrodes. \cite{Nogaret2010} This way, magnetic superlattices \cite{Carmona1995,Ye1995,Izawa1995}, open magnetic dots  \cite{Novoselov2002,Uzur2004}, magnetic stripes which generate electron transport via snake- and cycloid trajectories \cite{Nogaret2000}, and magnetic barriers \cite{Monzon1997,Johnson1997,Kubrak2000,Vancura2000,Cerchez2007,Tarasov2010} have been implemented. The measured conductance resonances could all be explained in terms of semiclassical trajectories, while suggestions of quantum states in magnetic field gradients without a classical equivalence \cite{Reijniers2002,Xu2007a} have remained unobserved.

Here, we report the observation of the quantum states predicted by Reijniers et al. [\onlinecite{Reijniers2002}] to exist at the zero point of a magnetic step from a negative to a positive magnetic field, oriented across a quantum wire. In the original proposal, these states are inaccessible to transport experiments due to diamagnetic shifts of the wire modes in the leads. We overcome this difficulty by using magnetic fields of finite extension in transport direction. After introducing the experimental setup and the sample characterization, we present the experimental results and interpret them with the help of numerical simulations.

\begin{figure}[h!]
\includegraphics[scale=1.0]{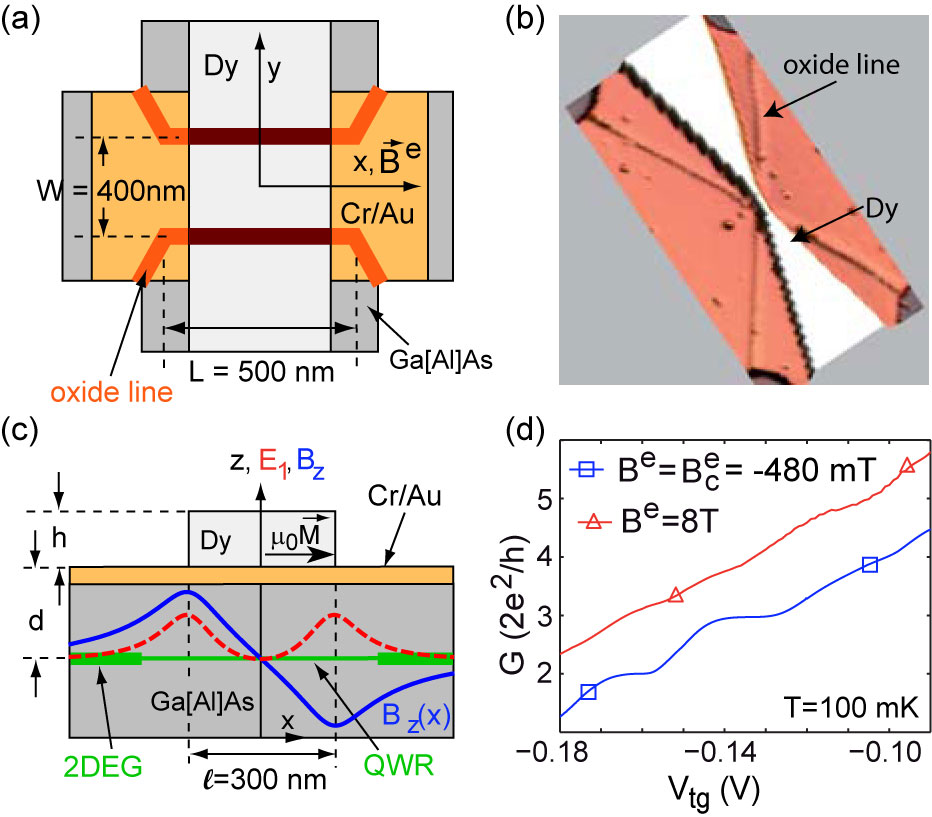}
\caption{(color online). (a) Scheme of the sample geometry. (b) Atomic force microscope picture of the Dy stripe on top of the oxide lines that define the QWR. (c) Schematic cross section of the structure in the xz - plane at y=0, with the magnetic field profile $B_z(x)$ (full blue line) and the energy of the first QWR mode $E_1(x)$ (dashed red line). (d) Conductance of the QWR as a function of the top gate for two applied magnetic fields $B^{e}$. }
\label{QS_IN_BGR_Fig1}
\end{figure}

A $\mathrm{GaAs/Al_xGa_{1-x}As}$ heterostructure with a two-dimensional electron gas (2DEG)  $45\,\mathrm{nm}$ below the surface is used. The 2DEG has an electron density of $3.8\times 10^{15}\,\mathrm{m^{-2}}$ and a mobility of $34\,\mathrm{m^2V^{-1}s^{-1}}$ at liquid helium temperatures. The sample layout is depicted in Fig. \ref{QS_IN_BGR_Fig1}(a) to (c). A Hall bar with ohmic contacts is patterned into the heterostructure by optical lithography, wet chemical etching and metallization, followed by an alloying step. The ballistic quantum wire (QWR) is defined by scanning probe lithography where the Hall bar surface is oxidized locally, which leads to depletion of the electron gas underneath. \cite{Held1999}  Its lithographic length $L$ and width $W$ are $L=500\,\mathrm{nm}$ and $W=400\,\mathrm{nm}$. The 2DEG to the sides of the QWR can be used as in-plane gates \cite{Wieck1990} to fine-tune the Fermi energy locally. The structure is covered by a homogeneous Cr/Au layer of $10\,\mathrm{nm}$ thickness, and a dysprosium (Dy) stripe of $\ell=300\,\mathrm{nm}$ width, oriented perpendicular to the wire, is defined on top by electron beam lithography. It has a height of $h\approx 250\,\mathrm{nm}$. An external magnetic field $\vec{B}^{e}=(B^{e}, 0,0)$ magnetizes the stripe along the x-direction. The z-component of the fringe field forms the desired magnetic field gradient $B_z(x)$ in transport direction, with a zero point at x=0.  For a magnetic dipole fringe field, $B_z(x)$ in the plane of the 2DEG is given by \cite{Vancura2000}

\begin{equation}
B_z(x) = \frac{\mu_0 M}{4\pi}\ln \left(\frac{A^{-}}{A^{+}}\right); A^{\pm}= \frac{(x\pm \ell/2)^2+d^2}{(x\pm \ell/2)^2+(d+h)^2}
\label{MDB_in_QWR_eq1}
\end{equation}

where $\mu_0M$ denotes the magnetization of the Dy stripe and $d$ is the distance of the 2DEG to the bottom of the Dy film. In Fig. \ref{QS_IN_BGR_Fig1}(c), $B_z(x)$ is plotted for our sample parameters: it corresponds to an approximately constant magnetic gradient in between the edges of the Dy stripe.  By Hall magnetometry, \cite{Monzon1997}, we estimate the  maximum stripe magnetization to $\mu_0 M\approx 2.1 \,\mathrm{T}$,\cite{Supplement} corresponding to a maximum value of $|B_z(\pm \ell/2)|=466\,\mathrm{mT}$ with a maximum gradient of $dB_z(x)/dx\approx 3\times 10^6\,\mathrm{T/m}$. Within a qualitative picture,  $B_z(x)$ predominantly influences QWR mode energies via a local diamagnetic shift, as sketched by the red dashed line in Fig. \ref{QS_IN_BGR_Fig1}(c). It is assumed that the x - component of the fringe field is irrelevant due to the strong electrostatic confinement in z-direction. Such two magnetic barriers in series of opposite polarity in a QWR have been discussed theoretically,\cite{Governale2000,Zhai2002,Papp2010,XuHengyi2011a} in particular as a tunable spin filter,\cite{Xu2001,Lu2002a,Seo2004,Jalil2005,Zhai2006} but has so far not been implemented experimentally, while experimental studies of magnetic double barriers in 2DEGs \cite{Kubrak1999,Bae2007} as well as of single magnetic barriers on QWRs \cite{Hugger2008,Tarasov2010} have been reported. We have observed the features reported below in two samples and focus below on the data from one sample. For comparative measurements, a second QWR of nominally identical geometry, but without the ferromagnetic stripe, was prepared on a different piece of the same heterostructure.

Measurements at temperatures $T\geq 2\,\mathrm{K}$ were performed in a $\mathrm{^4He}$ gas flow cryostat using standard lock-in techniques, while for $T<1\,\mathrm{K}$, a $\mathrm{^3He/^4He}$ dilution refrigerator was used. Both systems are equipped with a superconductive magnet and a rotating sample stage that allows to tune the QWR from perpendicular to parallel orientation with respect to $\vec{B}^{e}$. Parallel alignment is achieved by adjusting the Hall resistance to zero.

In Fig. \ref{QS_IN_BGR_Fig1} (d), the conductance of the QWR as a function of the top gate voltage at the coercive magnetic field $B^{e}_c = - 480\,\mathrm{mT}$ of the Dy stripe, is shown. We estimate the electron temperature to $T=100\,\mathrm{mK}$. Well-developed conductance plateaus with quantized values of $j\times 2e^2/h$ are observed for $j$ =2 and 3.\cite{Wees1988,Wharam1988} The plateau for $j=1$ (not shown) is weakly pronounced, which may be a consequence of the greatly decreased electron density in the 2DEG reservoirs, and the visibility of the j=4 plateau is low. In $B^{e}=8\,\mathrm{T}$, the conductance is shifted towards larger values. We attribute this shift to the magnetic barrier fields $B_z(x=\pm L/2)$, which suppresses backscattering at the entrance and the exit of the QWR. \cite{Houten1988} This appears plausible considering that our reference QWR without a Dy stripe on top shows, in the regime of six occupied modes at $B^{e}=0$, a positive magnetoconductance of about $2e^2/h$ as the homogeneous perpendicular magnetic field is increased from $0$ to $0.2\,\mathrm{T}$.\cite{Supplement} In addition, the plateaus tend to get suppressed, and the different, weak modulation of the conductance traces can not be assigned to the number of occupied modes. This is different as compared to homogeneous magnetic fields, which are known to increase the markedness of the conductance plateaus.\cite{Wees1988a} We can exclude spin splitting as possible origin, since measurements on the nominally identical QWR without a Dy stripe on top did not show signatures of spin splitting up to $B^{e}=10\,\mathrm{T}$, \cite{Supplement}  indicative of a small effective g factor in our QWRs, which may be due to screening by the homogeneous Cr/Au top gate. We therefore attribute the suppression of the conductance quantization and the emergence of additional features to the magnetic field profile $B_z(x)$.

\begin{figure}[ht]
\includegraphics[scale=1.0]{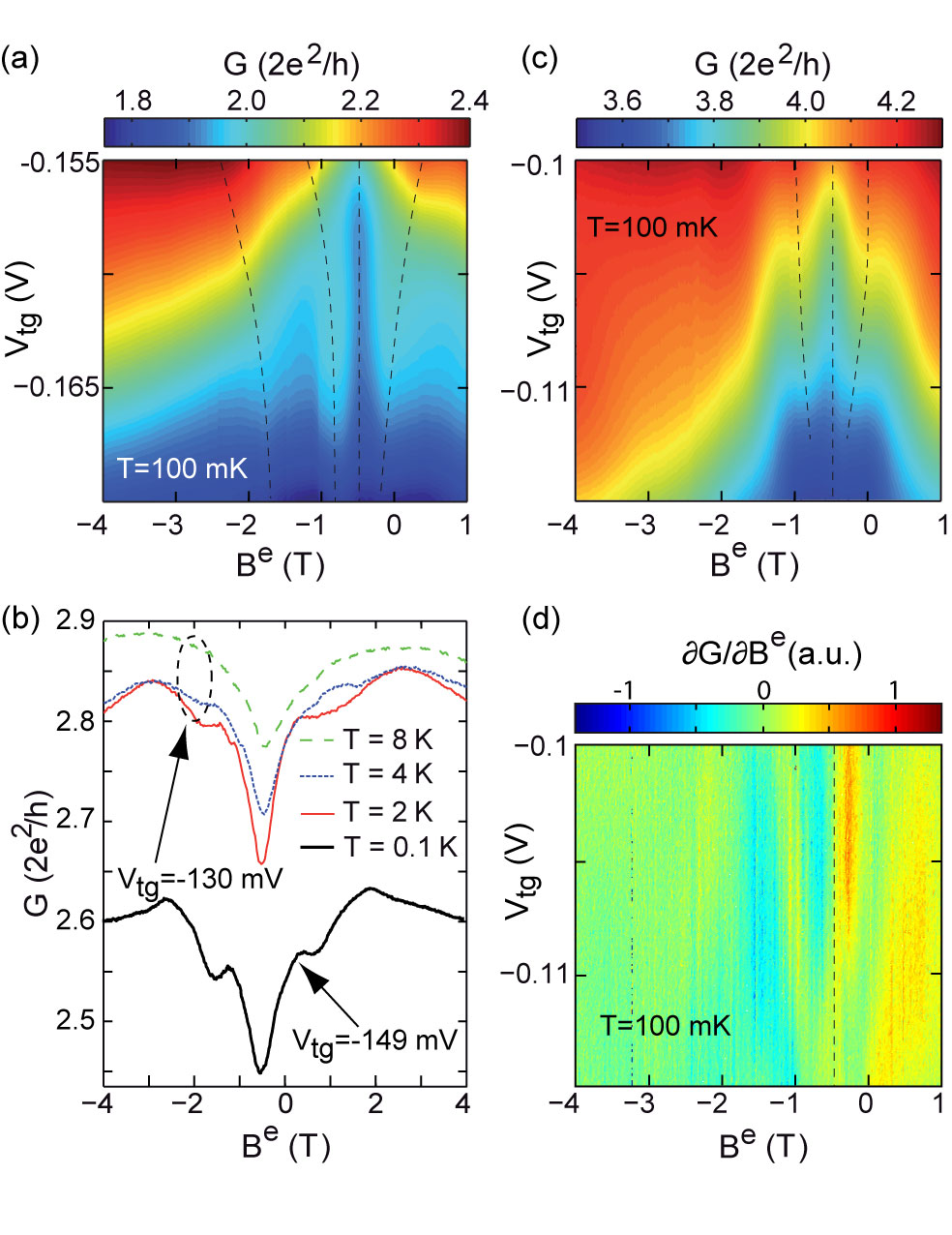}
\caption{(color online). Plots of $G(B^{e},V_{tg})$ around the plateaus 2 (a) and 4 (c). The vertical and bent dashed lines indicate the coercive magnetic field and some peak positions, respectively, as a guide to the eye. (b) Temperature dependence of $G(B^{e})$, as measured in two cryostats. (d) $\partial G(B^{e},V_{tg})/\partial B^{e}$ of the data shown in (c). }
\label{QS_IN_BGR_Fig2}
\end{figure}

Fig. \ref{QS_IN_BGR_Fig2} (a) shows the magnetoconductance of the QWR as a function of $B^{e}$ and of the top gate voltage $V_{tg}$ in the gate voltage interval where two modes are occupied. In all measurements shown, the voltage applied to the two in-plane gates was $-500\,\mathrm{mV}$. As a general trend, the conductance increases as $|B^{e}-B^{e}_c|$ is increased, which we attribute to the changing local magnetic field at the entrance and exit as mentioned above. Superimposed conductance peaks are visible at distances to $B^{e}_c$ which increase as $V_{tg}$ is increased, as indicated by the dashed lines. In a different cooldown of the same sample in the $\mathrm{^4He}$ gas flow cryostat, the same structures are observed, which disappear at a temperature of $T\approx 8\,\mathrm{K}$, see Fig. \ref{QS_IN_BGR_Fig2} (b). We can therefore exclude that the resonances are of classical origin.

The features are less pronounced around plateaus 3 to 7, as exemplified for the fourth plateau in Fig. \ref{QS_IN_BGR_Fig2} (c). In order to increase the visibility of resonances, we show in Fig. \ref{QS_IN_BGR_Fig2} (d) the numerical derivative $\partial G/\partial B^{e}$ of the data of (c). It can be seen that the resonance closest to $B_c�$ shifts from $|B^{e}-B^{e}_c|\approx 500\,\mathrm{mT}$ at $V_{tg}=-100\,\mathrm{mV}$ to $\approx 250\,\mathrm{mT}$ at $V_{tg}=-112\,\mathrm{mV}$. Furthermore, these features are observed as a function of both the top gate and the in-plane gate voltages,\cite{Supplement} indicating again that they originate from inside the QWR rather than from some scattering centers in the 2DEG.

We therefore attribute the conductance resonances to discrete states inside the QWR that form in the presence of the magnetic field profile $B_z(x)$. They are interpreted with the help of numerical simulations, which are based on a combination of the tight-binding model with recursive Green functions. \cite{Xu2007a,XuHengyi2011a} The parameters used in the simulation are adapted to our sample, with the exception that $L=3\,\mathrm{\mu m}$ and that due to numerical limitations, the quantum wire (electronic width $W=300\,\mathrm{nm}$) is directly attached to reservoirs, such that the transition from the QWR to the 2DEG is not captured. Since the oxide lines written by scanning probe lithography are known to generate superparabolic confinement, \cite{Fuhrer2001} hard walls are assumed. The wire is exposed to the profile $B_z(x)$ given by eq. \eqref{MDB_in_QWR_eq1} with a barrier spacing of $\ell=300\,\mathrm{nm}$ and an amplitude parameterized by the magnetization $\mu_0M$. All calculations have been carried out for zero temperature.

\begin{figure}[h!]
\includegraphics[scale=1.0]{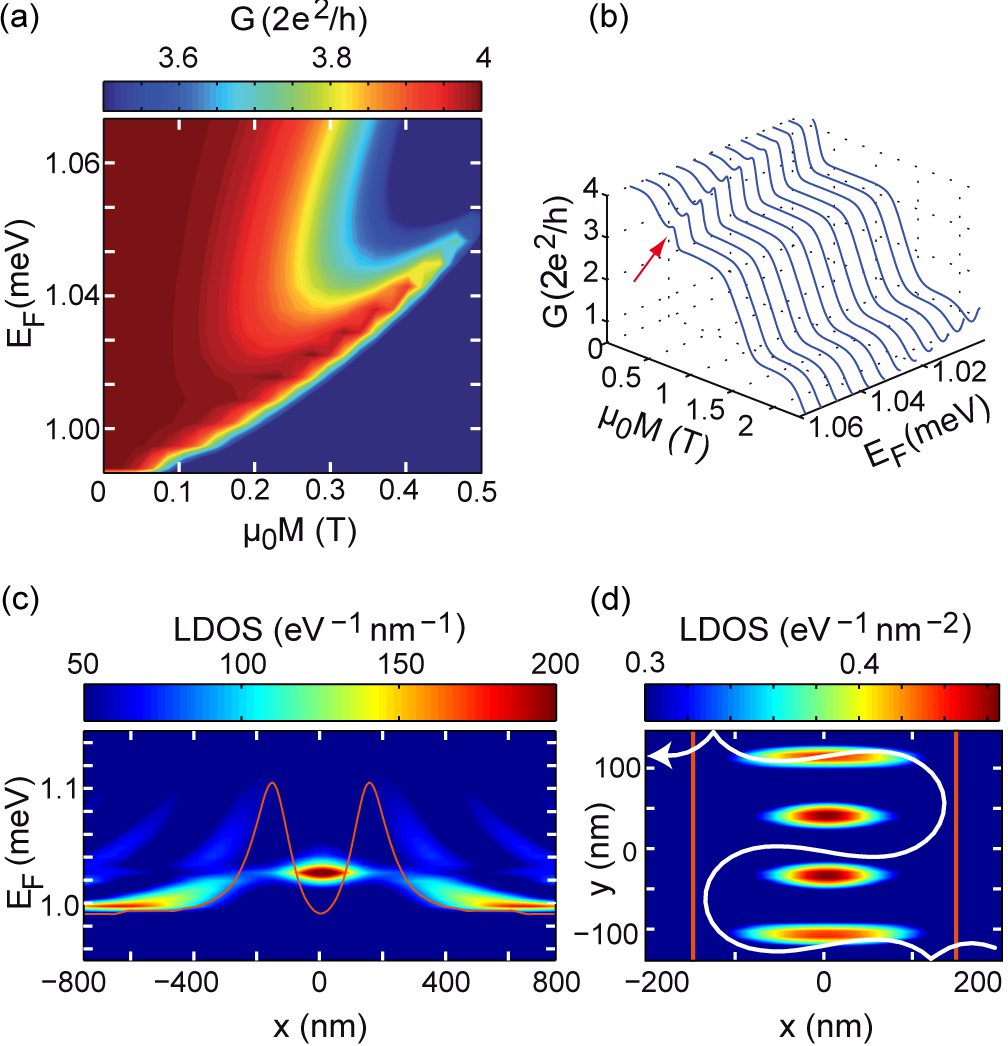}
\caption{(color online). Simulations for a realistic QWR exposed to a magnetic field gradient. The calculated conductance $G(E_{F}, \mu_{0}M)$ in the transition region between conductance plateau 3 and 4 (a) and for selected Fermi energies over a wider range of magnetizations (b), where the resonance of (a) is indicated by the arrow. (c) The LDOS along the QWR as a function of $E_{F}$, calculated for $\mu_{0} M=0.38\,\mathrm{T}$. One zero-dimensional state is visible at $x=0$. The red line is the energy of the fourth QWR mode. (d) $\mathrm{LDOS} (x,y)$ for the parameters of (c) at $E_{F}=1.04\,\mathrm{meV}$, and a classical electron trajectory (white line). The red lines indicate the extremal points of $B_z(x)$. }
\label{QS_IN_BGR_Fig3}
\end{figure}

We first focus on a clear-cut situation, as found at relatively large Fermi energies $E_F$ where four modes are occupied for $\mu_0M=0$, corresponding to Fig. \ref{QS_IN_BGR_Fig2} (c,d). In Fig. \ref{QS_IN_BGR_Fig3} (a), the calculated conductance as a function of $\mu_0M$ and $E_F$ is shown. The QWR modes get depopulated locally as either $E_F$ is decreased or $\mu_0M$ is increased. In the transition region between plateau 3 and 4, a conductance resonance is observed which shifts to larger magnetization as $E_F$ is increased, in qualitative agreement with the experimental results. Its amplitude depends sensitively on the parameters as well as on the conductance plateau considered, as illustrated by the calculated conductance over a larger range of magnetizations shown in Fig. \ref{QS_IN_BGR_Fig3} (b), where one resonance between plateau 3 and 4 is visible, but none between plateau 2 and 3.

The origin of such resonances are bound states at the zero point of $B_z(x)$ at $x=0$, which becomes apparent with the help of the local density of states (LDOS). In Fig. \ref{QS_IN_BGR_Fig3} (c), the LDOS as a function of x and the Fermi energy, integrated along the y-direction, is shown for a ferromagnetic stripe magnetization of $\mu_{0} M=0.38\,\mathrm{T}$, which is a plausible magnetization for $|B^{e}-B^{e}_c|=0.5\,\mathrm{T}$. \cite{Supplement} The x-dependent energy of the modes is represented by the red line. Its maxima correspond to the centers of the two magnetic barriers. A zero-dimensional state at $x=0$ emerges. The LDOS of this zero-dimensional state at $E_F=1.04\,\mathrm{meV}$ as a function of x and y is shown in Fig. \ref{QS_IN_BGR_Fig3} (d). It has the character of a standing wave in transverse direction, and the number of nodes indicates that it originates from the fourth wire mode.

\begin{figure}[h!]
\includegraphics[scale=1.0]{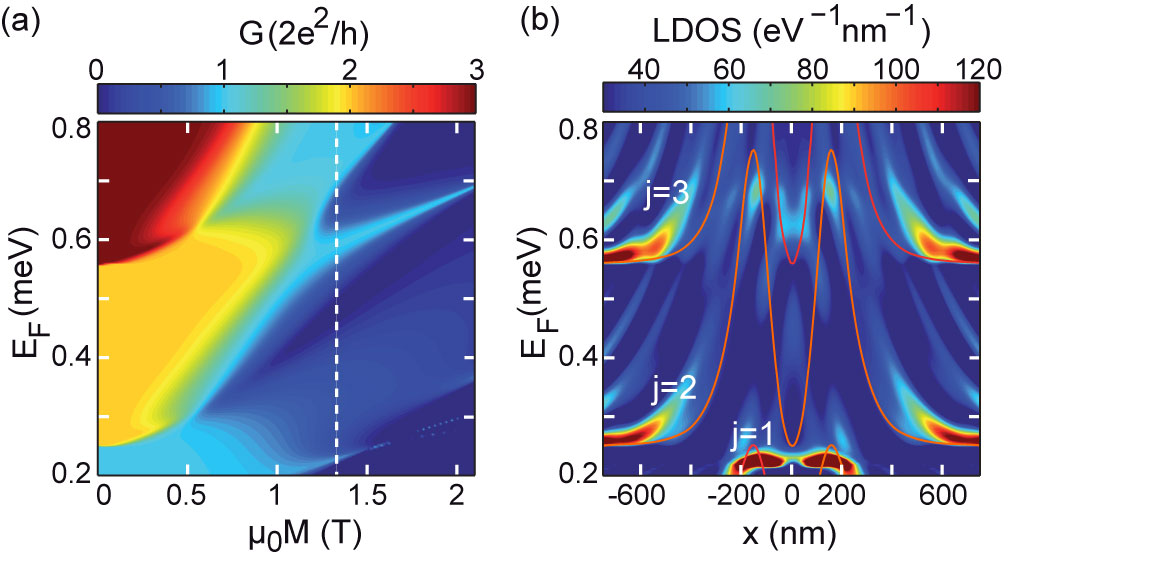}
\caption{(color online). Numerical studies of the system over a wide range of magnetizations and low Fermi energies. The calculated conductance is shown in (a). The dashed line marks the magnetization of $\mu_0M=1.33\,\mathrm{T}$ used for  the LDOS as a function of $E_F$ and the x-coordinate shown in (b). Here, j indicates the mode index.}
\label{QS_IN_BGR_Fig4}
\end{figure}

This state has the same structure as those residing at the zero point of an infinitely sharp magnetic step $B_z(x)=B_0(\Theta(x)-\frac{1}{2})$, studied theoretically by Reijniers et al. [\onlinecite{Reijniers2002}], where $\Theta(x)$ denotes the Heaviside step function and $B_0$ is the step height. As pointed out there, this state does not have a semiclassical equivalence, since classically, only a unidirectional snake type orbit exists at such a step. This is exemplified by one calculated trajectory, shown in Fig. \ref{QS_IN_BGR_Fig3} (d). A quantum calculation of the energy dispersion however shows that states do exist with a group velocity of opposite polarity to that one of the classical snake orbit, such that a standing wave can be formed. Our experimental implementation of this structure deviates in some respects from the infinitely sharp step. First of all, the sharp step corresponds to a constant diamagnetic shift along the QWR, except at $x=0$ where the shift is zero. Consequently, the effective confinement has the character of a $\delta$-potential and only one bound state per mode in the x-direction exists. However, in our implementation, the magnetic field gradient is finite and the effective confinement in x-direction is softer, such that several states along the x-direction become possible. Also, a striking paramagnetic shift of the bound states is found in Ref. [\onlinecite{Reijniers2002}], which reflects their increasing localization at the magnetic step as $B_0$ is increased. For our magnetic field profile, however, the effective potential becomes narrower as the magnetization increases, which causes the energies of the states to increase with increasing magnetization. Furthermore, we note that for the parameters considered here, all occupied modes contribute to the conductance at the Fermi level, but only the uppermost mode forms zero-dimensional states. We therefore expect that the transmission has the character of a Fano resonance,\cite{Fano1961,XuHengyi2011a} and that the LDOS contains admixtures of the propagating states belonging to lower lying modes. Within our implementation of the numerical algorithms, it is not possible to decompose the LDOS into contributions by different modes. A Fano character of the resonances should moreover also show up in their line shape. Due to the strongly varying background, however, such fits to our measured resonances are questionable.

The consequences of the deviations of our experimental implementation to the ideal magnetic step in a QWR are illustrated in Fig. \ref{QS_IN_BGR_Fig4}, where the conductance over a wide range of Fermi energies and magnetizations is shown. As the magnetization increases, the weakly pronounced conductance modulations evolve into sharp resonances, reflecting the increased effective confinement, see Fig. \ref{QS_IN_BGR_Fig4}(a). Furthermore, a second resonance per QWR mode can be formed. This can be seen, for example, in the conductance as a function of $\mu_0M$ for $E_F\approx 0.62\,\mathrm{meV}$, or as a function of $E_F$ at $\mu_0M= 1.33\,\mathrm{T}$. For the latter case, we show the LDOS ($E_F, x$) in Fig. \ref{QS_IN_BGR_Fig4}(b). Here, two LDOS maxima at $E_F \approx 0.62\,\mathrm{ meV}$ and $\approx 0.68 \,\mathrm{ meV}$ are visible, both belonging to the third QWR mode. Note that the bound state with larger energy shows a node in the x-direction at $x=0$. This energy structure reflects the facts that in our sample, the steepness of our magnetic step is finite and that the confinement in the x-direction is weaker than in the y-direction.

To summarize, we have observed resonant transmission through a ballistic quantum wire via hitherto unobserved magnetically bound states that reside at the zero point of a magnetic field gradient, as suggested by Reijniers et al. [\onlinecite{Reijniers2002}] These quantum states have no classical equivalence. We observe up to two such states per wire mode, all with a diamagnetic response. Both of these features are in contrast to the original proposal and we have explained them by the finite steepness of the magnetic step in our experimental implementation. The presence of several bound states per mode is also of relevance regarding an experimental realization of the suggested magnetic barrier spin filters, \cite{Lu2002a,Seo2004,Jalil2005,Zhai2006} since their energy spacing may be comparable to the spin splitting energy and thereby influence the achievable degree of spin polarization.\\

The authors would like to thank HHU D\"usseldorf for financial support and I. V. Zozoulenko for valuable discussions in relation to the Green function formalism. ADW is grateful to the DFH/UFA for support in the CDFA-05-06.

%\bibliography{LSHeinzel_2014_04_24}

%

\end{document}